# Ultracold Muonium Negative Ion Production

V. Dudnikov[1 a)] and A. Dudnikov[2]

[1]*Muons, Inc., 552 N. Batavia Ave, Batavia, IL 60510 USA;*
[2]*Budker Institute of Nuclear Physics, Novosibirsk, Russia*

[a)]*Corresponding author: Vadim@muonsinc.com*

A new, efficient method to produce ultracold negative muon ions is proposed. The muonium atom is made up of an antimuon and an electron and is given the chemical symbol Mu. A second electron with binding energy or electron affinity of 0.75 eV makes the Mu- ion, which is in many ways almost identical to the H- ion that is used for charge-exchange injection into most proton particle accelerators. Muonium negative ions were observed in 1987 by interactions of muons with a foil. Using the foil charge-exchange approach, the efficiency of transformation of muons to negative muonium ions has been very low $\sim 10^{-4}$. However, by using a hot tungsten or palladium single crystal foil or aerogel treated by cesium deposition, the production efficiency can be improved up to 50%. The process described here has surface muons focused onto a tungsten or palladium single crystal foil or aerogel (that can be heated up to 2000 Celsius) and partially covered by a cesium layer to provide a minimal work function. The negative muon ions can be extracted by a DC electric field and further accelerated by a linac and stripped in a thin foil. Charge exchange with a dense flow of positive or negative ions is proposed for conversion of slow muonium atoms into positive and negative muonium ions.

## INTRODUCTION

It has been more than 45 years since muon colliders and muon storage rings were proposed [1,2,3]. Interest in muon colliders increased significantly following the understanding that ionization cooling could be used to rapidly cool muon beams. Several workshops were held in the 1980s and 1990s, and in 1997 the Muon Collider Collaboration was formed, which later became the Neutrino Factory and Muon Collider Collaboration (NFMCC). By the late 1990's muon collider and neutrino factory design efforts were well-established worldwide. In 2007 the International Design Study for a Neutrino Factory (IDS-NF) was initiated. In 2011, muon R&D in the United States was consolidated into a single entity, the Muon Accelerator Program (MAP) [4,5]. In 2014, the P5 Committee lowered the priority for Muon Collider work, terminating further MAP funding [6]. The work described below represents continued interest by the Industrial Community, often supported in the past by SBIR and STTR grants [7], to develop new ideas for intense, cooled muon beams that are useful for colliders, neutrino factories, energy and intensity frontier experiments, as well as commercial applications such as a muon microscopy, cargo scanning and tomography. The muonium atom is made up of an antimuon and an electron and is given the chemical symbol Mu. A second electron with binding energy or electron affinity of 0.75 eV makes the Mu- ion, which is in many ways almost identical to the H- ion that is used for charge-exchange injection into most proton particle accelerators. Muonium negative ions were observed in 1987 [8,9] by interactions of muons with a foil. Using the foil charge-exchange approach, the efficiency of transformation of muons to negative muonium ions has been very low $\sim 10^{-4}$. However, using a hot tungsten or palladium single crystal foil or aerogel treated by cesium deposition, it can be improved up to 50%.

In the following sections we first describe the present technique for Mu+ production, then our proposed new approach that avoids the use of a complex laser by taking advantage of adding cesium to the foil and offers the prospect of increased production efficiency. Improvement estimates are made with comparisons to H- and positronium negative ion experience. Finally, we discuss possible locations and uses for testing the proposed approach. A new, efficient method to produce ultracold negative muon ions is proposed in [10].

## COLD MUONIUM NEGATIVE ION PRODUCTION

Ultraslow muons up to now have been generated by resonant ionization of thermal muonium atoms (Mu) generated from the surface of a hot tungsten foil placed at the end of an intense surface muon beam line. In order to efficiently ionize the Mu near the W surface, a resonant ionization scheme via the 1s-2p unbound transition has been used. The low emittance muon beam has been discussed in several scientific reports [11, 12, 13, 14]. A complex laser system has been used to efficiently ionize the Mu near the W or aerogel surface [11,12].

Cesiation is the addition of a small admixture of cesium to a gas discharge, increasing negative ion emission and decreasing electron emission below that of the negative ions [15,16,17,18,19]. Cesiation decreases the surface work function and increases the probability for back scattered and sputtered particles to escape as negative ions. It is difficult to control the surface work function during the discharge [20].

A positronium negative ion Ps- is a bound system consisting of a positron and two electrons. The binding energy of positronium is I= 6.9 eV. The binding energy of the additional electron in a positronium negative ion, the affinity, is S=0.32 eV [21].

Positronium negative ions are created when a positronium atom escaping from metal captures an electron from the surface. In regard to H-, the probability of formation of positronium negative ions strongly depends on the surface work function. When corrections beyond the third level were included for the first time, the Ps$^-$ decay rate was found to be 2.087963(12) ns$^{-1}$. In publication [22], a significant increase of positronium negative ion emission was observed after deposition of cesium on the surface of a single tungsten crystal. In [23] it was proposed to use this effect for control of the surface work function in surface plasma sources.

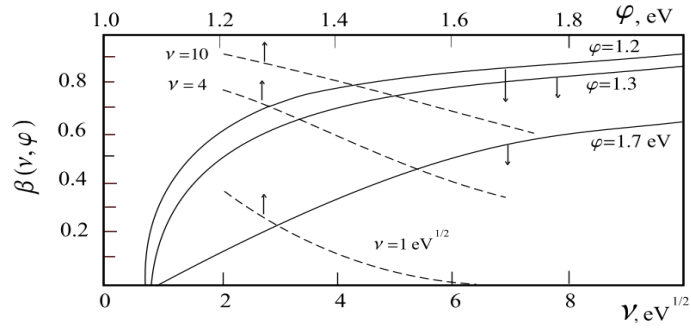

**FIGURE 1**. Calculated probability of sputtered and reflected particles escaping as H- as a function of work function and escape speed.

As shown by Kishinevskii [24,25], when atoms approach the metal surface, the electron affinity level goes down and widens. If the surface work function is not significantly greater than the electron affinity, at some distance from the surface, the electron affinity becomes lower than the Fermi level and an electron can jump from the metal into the electron affinity level. If such a negative ion moves fast enough away from the surface, the additional electron cannot tunnel back with any significant probability. This probability was calculated by Kishinevskii [24,25].

The probability of negative ion neutralization as a function of the distance R from the surface is (in atomic units):

$$W(R) = B^2 \gamma \Gamma^2 (1-\lambda) e^{-\frac{1}{2\gamma}} \left(\frac{R}{2\lambda}\right)^{2\lambda} e^{-2\gamma R}$$

Where $\gamma = 0.236, \frac{\gamma^2}{2} = S$ is the electron affinity in atomic units,

$\lambda = \frac{1}{4\sqrt{2\varphi}}$, $\varphi$ is the work function, $B = 1.68$ $is$ the coefficient in the wave function of the electron in a hydrogen negative ion far away from the surface $\Psi \approx \left(\frac{B\sqrt{2\gamma}}{\sqrt{4\pi}}\right)\left(\frac{e^{\gamma r}}{r}\right)$, $\Gamma$ is the gamma function.

The ionization coefficient is

$$\beta^- = exp - \int_{R^0}^{\infty} W(R) \frac{dR}{v_\perp(R)} \approx exp - \frac{B^2 \Gamma^2 (1-\lambda)}{2 v_\perp R_0} \sqrt{\frac{\varphi}{U_0}} e^{-\frac{1}{2\gamma}} \left(\frac{R_0}{2\lambda}\right)^{2\lambda} e^{-2\gamma R_0}$$

where $R_0 = \frac{1}{4(\varphi-S)}$, $U_0$ is the depth of the potential well in the metal, $v_\perp(R)$ is the speed of the negative ion escaping from the surface at a point of $R_0$. For Mu- the ionization coefficient should be higher than for H- because its speed is higher for the same escaping energy.

Figure 1 shows the calculated probability of sputtered and reflected particles escaping as H⁻. For a realistic work function $\varphi > 1.7$ eV, the probability, $\beta^-$, of a negative ion forming on the metal surface is proportional to the escape velocity of the ejected ion v (transverse to the surface) and inversely proportional to the surface work function less the affinity, as shown in Fig. 1 for low velocities; at larger velocities, it saturates.

$$\beta^- = 0.12\ (v-v_o)/(\varphi-S), \text{ where } v_o = (\varphi-S)^{1/2} \text{ in } (eV)^{1/2}, (\varphi-S) \text{ is in eV.}$$

Figure 2 shows the dependence of the H⁻ production on the work function of the Mo surface in cesiated hydrogen discharge for different bias voltages (escaping velocities) [20], as shown in Figure 1.

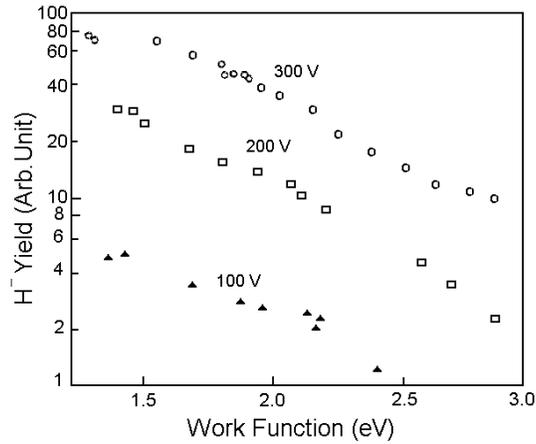

**FIG. 2.** Dependence of H⁻ production on the work function of the Mo surface in cesiated hydrogen discharge for different bias voltages [20].

To find the total ionization coefficient we must integrate $\beta^-$ (v) with the distribution function of the ejected particles with respect to the velocity $v(R_0)$. The integration should be carried out over all particles whose kinetic energy exceeds $\varphi-S$ in the perpendicular direction. This is the energy which a negative ion must have in order to overcome the attraction toward the surface by image forces and to depart from the distance $R_0$ to infinity.

For production of muonium negative ions, a proton (deuteron) beam is first injected into a primary pion production target made of 20-30 mm thick, disc-shaped, isotropic graphite. About 5% of the proton beam is consumed in the target. Most positive muon beams are generated from pions stopped at the inner surface layer of the primary production target and decaying at rest, hence the common name, surface muons. The muon is emitted isotropically from the pion with momentum 29.8 MeV/c and kinetic energy 4.119 MeV (in the rest frame of the pion). The intensity of the surface muon beam can be estimated from the number of pions stopped near to the surface. The extraction angle of two beam transport lines is 60 degrees relative to the proton beamline (forward) direction. The required acceptance of the beam transport line is evaluated to be about 100 msr, taking into account the extraction angle for an effective surface-muon-emission rate of 15 000 muons/s. The transported muon beam will be focused onto a palladium single crystal foil target or to aerogel to produce muonium negative ions. The transmission efficiency of the beamline is preferred to be as high as possible. The focused beam spot size at the foil is required to be less than 4 cm in diameter. Achieving the smallest beam spot size increases the slow muon beam intensity. An example of negative muonium ion acceleration is presented in [26].

# MUONIUM NEGATIVE ION PRODUCTION TARGET MADE OF SINGLE CRYSTAL WITH DEPOSITED CESIUM LAYER

The 29.8 MeV/c muon beam will be focused onto the palladium target or aerogel to produce muonium (Mu), which is an exotic atom made up μ+ and electron. The muonium negative ion is formed by electron capture of Mu near the surface of the hot palladium foil or aerogel with cesium. Then the muonium negative ions can be evaporated to the vacuum with thermal velocity. By producing muonium, the μ+ beam will be effectively stopped, yet maintain its polarization. The electrons are then stripped from the muonium negative ions using a thin foil. The ultracold muons produced this way will be fully polarized, with small transverse momentum.

In Yoshida et al. [11] and Miyake et al. [12] ultraslow muons are generated by resonant ionization of thermal Mu atoms generated from the surface of a hot tungsten foil placed at the intense surface muon beam line. In order to efficiently ionize the Mu near the W surface, a resonant ionization scheme is adopted via the {1S-2P-unbound} transition. In order to induce 1S-2P transitions, Lyman-α light of 122.088 nm is needed. To generate this Lyman-α (VUV, Vacuum Ultra Violet) laser light, the authors adopted resonant four-wave frequency mixing ($\omega_{VUV} = 2\omega_r - \omega_t$) where two 212.5 nm photons($\omega_r$) are used for two-photon resonant excitation of the $4P^5 5P[5/2]$ state in Kr, subtracted by a photon with a tunable difference wavelength ($\omega_t$ ; 820 nm). In new laser system developed by the A04 group (Wada et. al.), $\omega_r$ (100 mJ/p, 2ns) is generated as the 5th harmonic of 1062 nm (1 J/p), which consists of an all solid state laser system, such as diode laser, fiber amplifier, regeneration amplifier, OPA, OPG etc., generating expectedly more than 71μJ/cm$^2$ Lyman-α light of the saturation intensity for the Doppler broadened Mu at 2000 K.

The muon beam is created by an 8 GeV proton beam striking a production target and a system of superconducting solenoids that efficiently collect pions and transport their daughter muons to a stopping target. J-PARC was funded to install a second muon beamline, called the U-Line, which consists of a large acceptance solenoid made of mineral insulation cables (MIC), a superconducting curved transport solenoid magnet and a superconducting axial focusing magnet system [11,12]. There, it is possible to collect surface muons with a large acceptance of 400 mSr. Compared to the conventional beamlines such as D-Line, the large acceptance of the front-end solenoid will allow for the capture of more than 10 times the intensity of pulsed muons [11]. With a muon capture of $5 \times 10^8$/s surface muons, can be collected $2 \times 10^8$/s surface muons on the W target in the Mu chamber, with an approximate transport efficiency of 40 %.

A schematic of our device for efficient production of slow muonium negative ions is shown in Fig. 3. The main component of this system is a single crystal of tungsten or palladium or aerogel with cesium deposition and an extraction system. The work function can be controlled by photo effect registration.

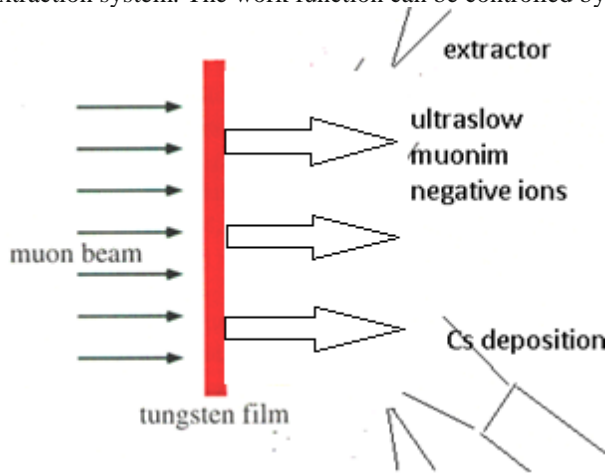

**FIGURE** 3. Schematic diagram of efficient ultraslow muonium negative ion production.

The target should have a possibility to be flashed up to 2000 C. Enclose should be heated for outgassing.
Mu mesons can be converted into muonium negative ions when they hit the thick palladium single crystal with the half of monolayer of cesium. The negative muonium ions escaping the crystal can be accelerated up to 30-50 keV and directed to the stripping foil and accelerated again as muons or can be accelerated without stripping to the injection energy of the g-2 experiment and charge exchange injected (with laser stripping) into the ring.

# IONIZATION OF MUONIUM ATOMS BY RESONANT CHARGE EXCHANGE WITH SLOW IONS

Another possibility of efficient ionization of muonium atoms is resonance charge exchange with slow ions. The cross section of this process is very high.

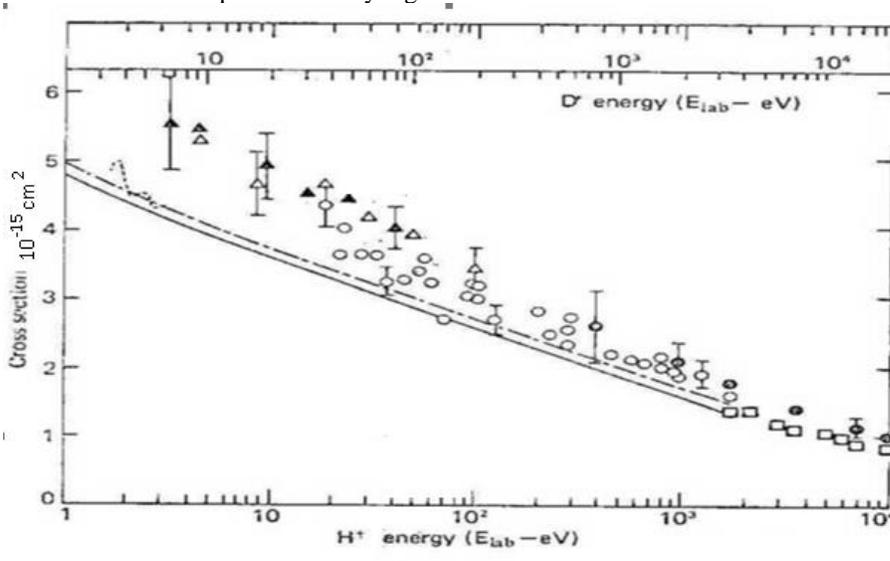

**FIGURE** 4. Cross section of resonance charge exchange with protons.

The cross section of resonance charge exchange with protons shown in Fig. 4 is reaches $5 \cdot 10^{-15} cm^2$ at 10 eV energy. The cross section of resonance charge exchange with negative hydrogen ions, shown in Fig. 5, reaches $10^{-14}$ $cm^2$ at 10eV energy.

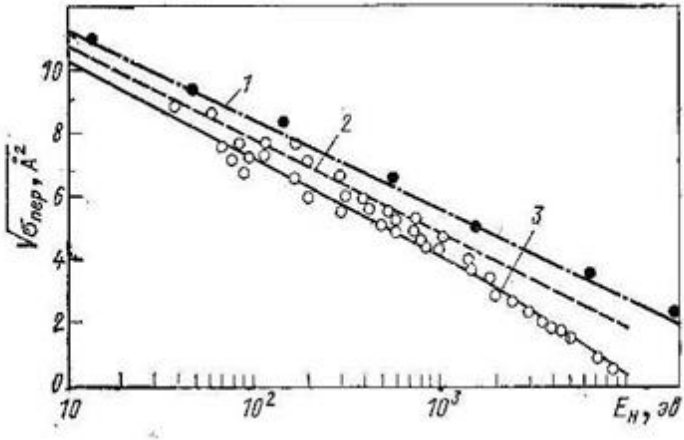

**FIGURE** 5. The cross section of resonance charge exchange with negative hydrogen ions.

These processes were successfully used for highly efficient ionization of polarized hydrogen and deuterium atoms [27,28,29,30]. For generation of highly dense plasma an arc discharge plasma source with diaphragmed channel developed in BINP [31,32], was used. A schematic of this plasma source is shown in Fig. 6. It consist of a pulsed gas valve 1, triggering electrode 2, cathode 3, barrier diaphragm 4, washed channel 5 and anode 6. An arc discharge with a current up to 300 A is triggered in the washed channel by triggering electrode 2. A dense plasma flux is extended through an anode channel. The muonium cooling target 7 accepts the muons flux and produces cold muonium atoms. The muonium atoms leaving the muonium cooling target are ionized by resonance charge exchange with plasma protons and move to the plasma grid 8 by the voltage between the cooling target and the plasma grid. The extraction grid 9 extracts ionized muons and accelerates them as a flux of accelerated cold muons 11. The efficiency of polarized positive ion production is up to 40%.

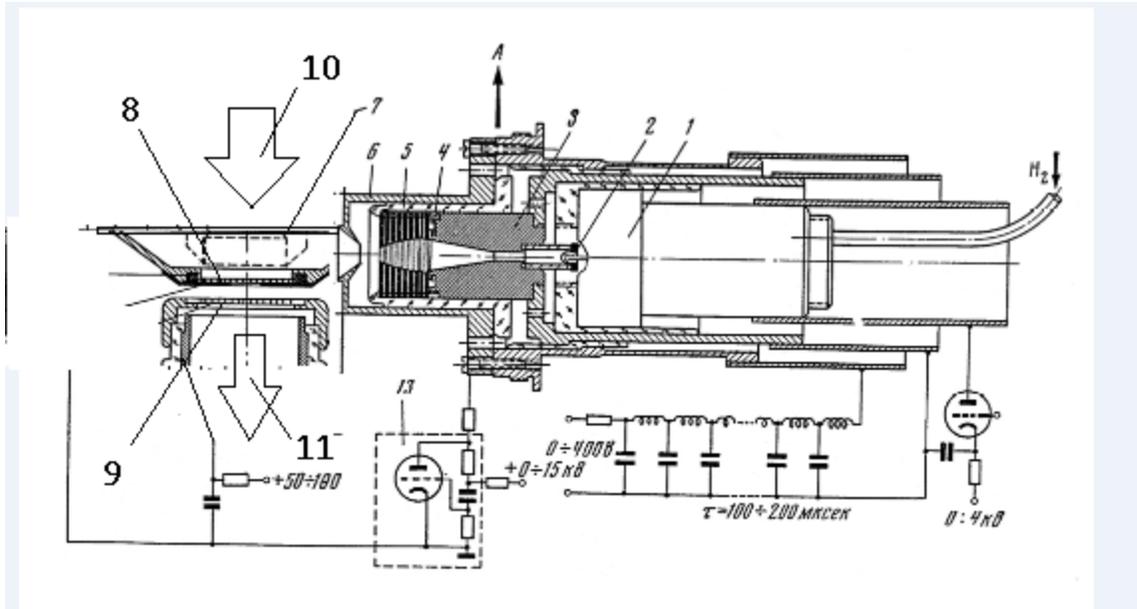

**FIGURE** 6. Schematic of plasma source and ionization of flux of muonium atoms.
1- Pulsed gas valve, 2- triggering electrode, 3- cathode, 4-barier diaphragm, 5- washed channel, 6-anode, 7- muonium cooling target, 8- plasma grid, 9- extraction grid, 10-flux of hot muons, 11-flux of accelerated cold muons.

The flux of protons from the arc discharge plasma source can be efficiently converted to the flux of hydrogen negative ions in the surface plasma ionizer shown in Fig. 7.

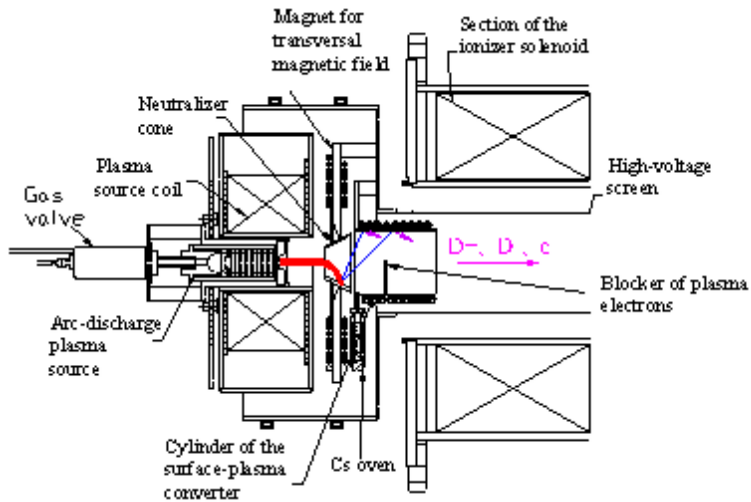

**FIGURE** 7. Surface plasma ionizer for conversion of the proton flux of arc discharge plasma source into the flux of negative hydrogen ions.

This flux of negative ions can be used for resonance charge exchange with muonium atoms for efficient production of negative muonium ions as polarized negative ion were produced in works [27,2]. The efficiency of polarized negative ion production was up to 12%.

Another possibility of polarized atom ionization was proposed in [33]. The schematic of this device is shown in Fig. 8. Plasma sources for intense negative ion production rely on the efficient conversion of positive plasma ions to negative ion species on surfaces with reduced work functions [15,16]. The emission is enhanced by using cesium to lower the work function of the emission surface, a well-known and often-used technique. Shaping of the emission surface into concave spherical surfaces, known as geometrical focusing or self-extraction [34,14], results in a natural focusing of the negative ions that increases the negative ion current density. This has demonstrated an increase in negative ion emission even on cesiated surfaces [35]. The concept is illustrated in Figure 9.

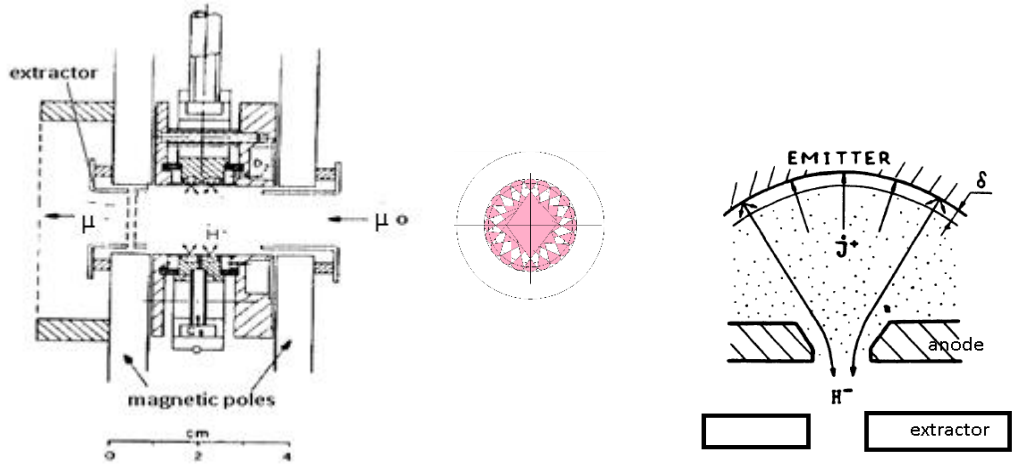

**FIGURE** 8: Surface plasma source-based ionizer for resonant charge-exchange ionization of muonium atoms to extract the muonium negative ions.
**FIGURE** 9: Cylindrical or spherical shaping of the emitter surface geometrically focuses the emitted negative ions.

The concave spherical emitter electrodes in the surface plasma source can be arranged such that the generated negative hydrogen ions are focused onto apertures leading away from the discharge, into a charge exchange area. The ionizer combines the surface plasma source with a short charge-exchange region into which the unpolarized H- ions are injected radially. The muonium atoms will be injected on axis into one end of this charge-exchange region with the extraction grid at the opposite end. The concept is illustrated in Figure 8.